\documentclass[%
 reprint,
 amsmath,amssymb,
 aps,
]{revtex4-2}

\usepackage{graphicx}% Include figure files
\usepackage{booktabs} 
\usepackage{dcolumn}% Align table columns on decimal point
\usepackage{bm}% bold math
\usepackage{xcolor} 
\usepackage[utf8]{inputenc}
\usepackage{graphicx,amsmath,amssymb}
\usepackage[colorlinks]{hyperref}

\makeatletter
\newcommand{\beginsupplement}{%
  \setcounter{table}{0}\renewcommand{\thetable}{S\arabic{table}}%
  \setcounter{figure}{0}\renewcommand{\thefigure}{S\arabic{figure}}%
  \setcounter{equation}{0}\renewcommand{\theequation}{S\arabic{equation}}%
  \setcounter{section}{0}\renewcommand{\thesection}{S\arabic{section}}%
}

\renewcommand\@biblabel[1]{[S#1]}
\makeatother

\begin{document}

%\preprint{APS/123-QED}

\title{Strain-Induced Antiferromagnetic-to-Altermagnetic Phase Transition\\ and Topology in $(\mathrm{CrO}_2)_1/(\mathrm{TaO}_2)_2$ Superlattice}% Force line breaks with \\

\author{Wanfei Shan}
\email{wfshan3132@ucla.edu}
 \affiliation{Division of Physical Sciences, College of Letters and Science, University of California, Los Angeles, 90095, California, USA.}%Lines break automatically or can be forced with \\
\author{Qun Yang}%
\affiliation{Division of Physical Sciences, College of Letters and Science, University of California, Los Angeles, 90095, California, USA.}%Lines break automatically or can be forced with \\
\author{Prineha Narang}
\email{prineha@ucla.edu}
\affiliation{Division of Physical Sciences, College of Letters and Science, University of California, Los Angeles, 90095, California, USA}
\affiliation{Department of Electrical and Computer Engineering, University of California, Los Angeles, 90095, California, USA.}

\begin{abstract}
Topological aspects in altermagnets have come into focus recently, and tuning the antiferromagnetic (AFM) state into an altermagnetic phase remains an active frontier. We realize both within a rutile superlattice here in this paper. With first principles calculation, we show that a uniaxial strain of only 0.5$\%$ along the c axis converts the $(\mathrm{CrO}_2)_1/(\mathrm{TaO}_2)_2$ rutile superlattice from a trivial antiferromagnet into an altermagnet with topology accompanied by a weak SOC. The strain opens a spin-dependent band splitting of $\sim 1.1 eV$ and, despite the weak SOC together with in-plane magnetic moment orientation, generates an intrinsic anomalous Hall conductivity of order $10^3 S/cm$, comparable magnitude to that in ferromagnetic Weyl semimetals. Tiny SOC here with in-plane \(\text{N\'eel}\) orientation gaps out the Weyl nodal rings, giving rise to 16 Weyl points in the superlattice. Thus, we point out a simple route toward strain and field tunable, low-dissipation altermagnetic electronics.
\end{abstract}
\maketitle

Altermagnets were introduced as a class of magnetic materials in 2021 first in theoretical predictions \cite{PhysRevX.12.040501, PhysRevX.12.031042} followed by experimental confirmation in 2024 \cite{s41586-023-06907-7}. Unlike conventional antiferromagnets, the magnetic sublattices in altermagnets are related by rotations or mirror translation followed by time-reversion symmetry $\mathcal{T}$, rather than by inversion or translation operations. This distinct symmetry relationship gives rise to non-relativistic spin splitting along certain high-symmetry directions in momentum space, despite a vanishing net magnetization. Thus, alter magnets exhibit a rare combination of ferromagnetic (FM) such as spin splitting, antiferromagnetic (AFM) stability, and symmetry-tuned responses, making them ideal for high-density magnetic memories in spintronics and magneto-optical optoelectronic devices \cite{s41578-025-00779-1, sun2025symmetrybreakingmagnetoopticaleffectsaltermagnets}. \par
In addition to their unique magnetic signatures, altermagnets have emerged as a fertile ground for realizing functional quantum phases. The breaking of the combined time-reversal and spatial-inversion symmetry ($\mathcal{PT}$), along with crystallographic symmetries, makes altermagnets a promising host for topological phases and anomalous Hall effect under certain \(\text{N\'eel}\) vector orientations \cite{10.1038/s41928-022-00866-z}. These robust topological states hold promise for dissipationless electronics and quantum information technologies. However, experimentally verified topological altermagnets are extremely rare \cite{li2024topologicalweylaltermagnetismcrsb, acs.nanolett.5c00482}. Theoretical efforts have begun to identify candidate materials in both two and three dimensions \cite{PhysRevLett.134.096703}, but the field remains in its early stages. In particular, the realization of topological phase transitions in altermagnetic systems through experimentally accessible control parameters remains largely unexplored and offers promising opportunities for both fundamental understanding and device applications.\par
In this work, we propose the $(\mathrm{CrO}_2)_1/(\mathrm{TaO}_2)_2$ superlattice as a viable candidate hosting symmetry-protected topological states that bridge this gap. The $(\mathrm{CrO}_2)_1/(\mathrm{TaO}_2)_2$ superlattice undergoes a strain-tunable AFM-to-altermagnetic phase transition with a remarkably uniaxial strain of merely 0.5$\%$. The spin–resolved bands acquire a splitting of $\sim1.09 eV$, comparable to that reported for RuO$_2$ \cite{PhysRevX.12.031042}. Notably, this topological phase emerges despite the intrinsic weak spin–orbit coupling (SOC), which gaps out the Weyl nodal lines in this altermagnet and results in 16 Weyl points. These results outline a practical route to engineering barely explored topological features in altermagnets and broaden the landscape of candidates for spintronics and quantum technologies.\par

\begin{figure}
    \centering
    \includegraphics[width=1\linewidth]{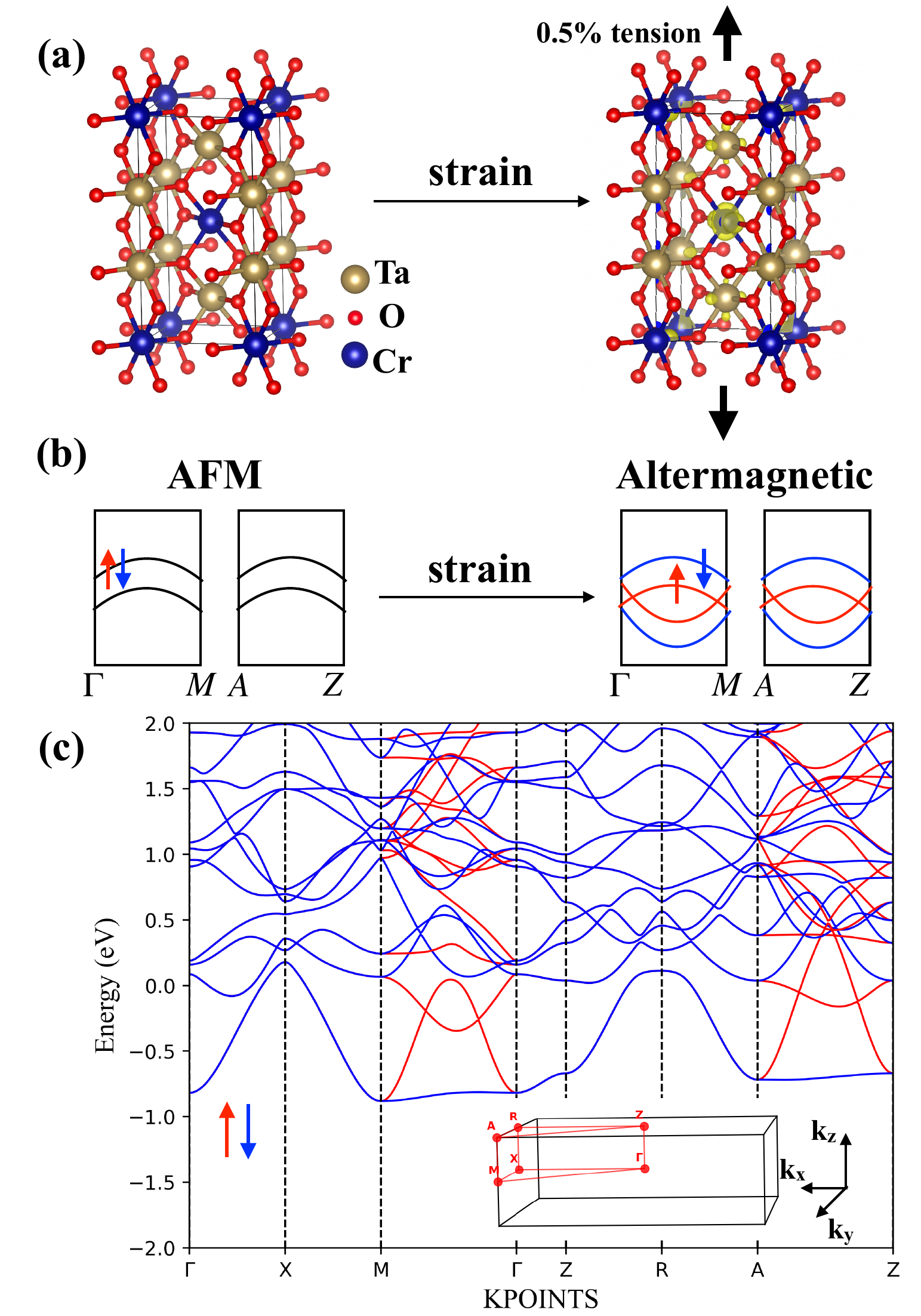}
    \caption{Electronic structure of the $(\mathrm{CrO}_2)_1/(\mathrm{TaO}_2)_2$ superlattice. (a) describes the $(\mathrm{CrO}_2)_1/(\mathrm{TaO}_2)_2$ superlattice from intrinsic superlattice to the one under tension; and isosurface plot of the partial charge density integrated within the energy window from $–1$ eV up to the Fermi level is shown with external strain. Blue, red, and dark yellow spheres denote Cr, O, and Ta atoms, respectively. (b) exhibits the magnetic phase transition under strain, especially along $\Gamma-M$ and A-Z direction. (c) Electronic band structure of the superlattice in its altermagnetic ground state, with red and blue lines corresponding to the spin-up and spin-down channels, respectively. High-symmetry directions used in the band structure plot, as indicated in the Brillouin zone (BZ) in the inset at the bottom right.}
    \label{fig: electronic structure}
\end{figure}

Both CrO$_2$ and TaO$_2$ crystallize in the space group P4$_2$/\textit{mnm} at room temperature, as illustrated in Fig.~\ref{fig: electronic structure}(a). For the rutile structure, the symmorphic operations contain inversion $I$, twofold rotation C$_2$ with the [100], [010], and [110] being the axes, fourfold rotation C$_4$ axis along [001], and mirror reflection M$_z$. In addition, nonsymmorphic operations are combining the mirror reflection, rotation and a non-symmorphic shift, such as \{$M_x|\tau$\} and \{$C_{4z}|\tau$\} with $\tau=(\frac{1}{2}, \frac{1}{2}, \frac{1}{2})$.\par
The $(\mathrm{CrO}_2)_1/(\mathrm{TaO}_2)_2$ superlattice retains the rutile structure under a uniaxial tensile strain of $0.5\%$, stabilizing in an altermagnetic ground state after structural relaxation. Each Cr atom carries a magnetic moment of approximately 3.3 $\mu_B$ by using both GGA+U (U$_{eff}$= 2, and 3 eV) \cite{10.1063/1.481183, 10.1088/1361-648X/aca19a, Shan_2023} and strongly constrained and appropriately normed (SCAN) plus a minimal U (U$_{eff}$= 0.2 eV) method, which is significantly larger than that calculated in RuO$_2$. Recent experiments on the prototypical altermagnet RuO$_2$ have revealed negligible spin splitting, which is attributed to the overestimation of the Hubbard U in initial theoretical predictions \cite{PhysRevLett.133.176401, PhysRevLett.132.166702}. In stark contrast, we demonstrate that the altermagnetic state in our superlattice is remarkably robust, exhibiting substantial spin splitting even for a minimal $U{\mathrm{_eff}} = 2$ eV. The energy splitting related by \{$C_{4z}|\tau$\} translation between the two Cr atoms in the superlattice only happens along $\Gamma$ to $M$ and $A$ to $Z$ directions in the band structure in Fig.~\ref{fig: electronic structure}(b). The largest energy split is calculated by \(\Delta_\epsilon=\epsilon_{n,k,\uparrow}-\epsilon_{n,k,\downarrow}\sim 1.09 eV\), the same magnitude as RuO$_2$ \cite{PhysRevX.12.031042}. It reflects the relatively strong exchange interaction between two sublattices, which is substantially larger than that in MnF$_2$ originated from weak higher-order anisotropic exchange interactions \cite{PhysRevLett.134.226702}.\par
In the absence of strain, the $(\mathrm{CrO}_2)_1/(\mathrm{TaO}_2)_2$ superlattice is a trivial AFM insulator, consistent with previous calculations \cite{Shan_2023}. Applying uniaxial strain induces interfacial charge redistribution with the partial charge density in the energy range $[-1,0]$ eV mainly localized on Cr atoms, in sharp contrast to the unstrained case, and triggers an insulator-to-metal transition as evidenced by the shifted density of states (see Supplementary Material). This charge transfer coincides with a transition from AFM to altermagnetic, which remains robust under moderate electron doping, further confirming the stability of the strain-induced altermagnetic state. \par
\begin{figure*}
    \centering
    \includegraphics[width=1\linewidth]{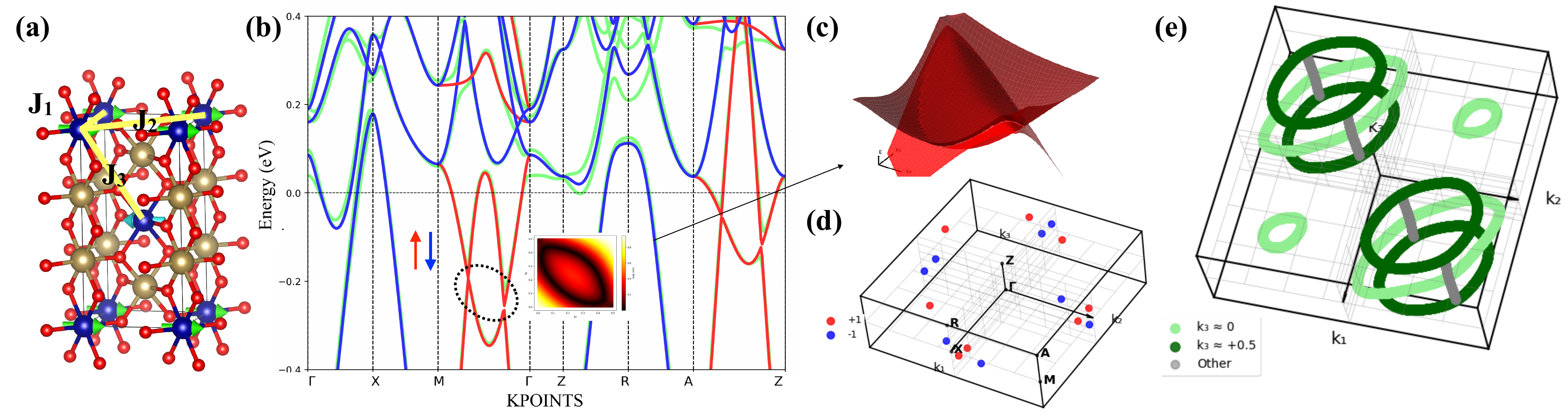}
    \caption{The crystal structure of the superlattice $(\mathrm{CrO}_2)_1/(\mathrm{TaO}_2)_2$ in altermagnetic phase, with local magnetic moments indicated, is shown in (a). (b) The electronic band structure of the altermagnetic phase, with red and blue lines corresponding to the spin-up and spin-down channels, respectively. The light green lines depict the band structure with spin aligned along the $x$-direction, incorporating spin–orbit coupling (SOC). (c) describes the three-dimensional band structure in the $k_z=0$ plane without SOC, corresponding to the red bands in (b). The Weyl points in the BZ are shown in (d). The distribution of nodal lines for the spin-up channel in the BZ is shown in (e), with light green indicating the ring in the $k_z=0$ plane, dark green indicating the ring in the $k_z=0.5$ plane, and gray lines indicating the lines between the $k_z=\pm0.5$ plane.}
    \label{fig:AHE}
\end{figure*}

To describe the exchange interaction in this material according to the relatively small magnetic anisotropy energy (MAE) of 0.00507 meV/per unit cell, we use the semi-classical Heisenberg model: 
\begin{equation}
    H=\sum_{i,j}J_{ij} \mathbf{S_i}\cdot \mathbf{S_j}
\end{equation}
where $J_{ij}$ represents the exchange interaction parameters between the spins, marked in Fig.~\ref{fig:AHE}(a). By mapping the energies of eight different AFM and FM configurations onto this formula, we obtain \(J_1=-0.328 meV\), \(J_2=-2.923 meV\), and \(J_3=0.180 meV\). These values indicate that the magnetic moments tend to align parallel within the magnetic atomic plane and antiparallel between atomic planes. Notably, the exchange interaction parameter, $J_2$, associated with the observed spin splitting, gives rise to the substantial band splitting near the Fermi level. Consequently, the $(\mathrm{CrO}_2)_1/(\mathrm{TaO}_2)_2$ superlattice emerges as a viable altermagnetic system, realizable under experimentally achievable strain, with spin splitting likely observable using spin- and angle-resolved photoemission spectroscopy (SARPES). \par
\begin{figure}
    \centering
    \includegraphics[width=1\linewidth]{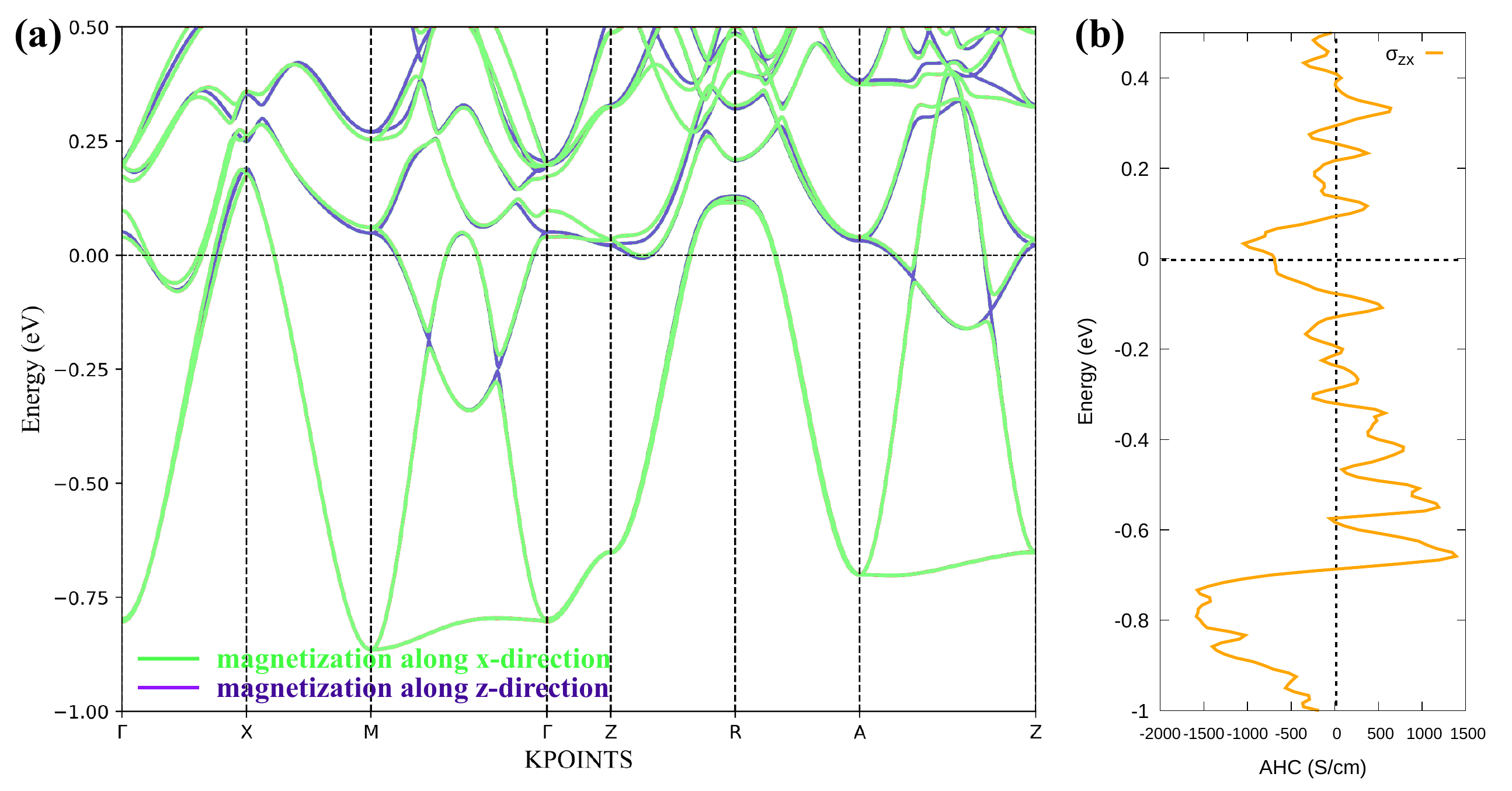}
    \caption{Electronic band structure of the $(\mathrm{CrO}_2)_1/(\mathrm{TaO}_2)_2$ superlattice with spin aligned along $z$-direction (violet) and $x$-direction (light green), including spin–orbit coupling (SOC), is shown in (a). The Anomalous Hall Conductivity (S/cm) for spin orientation along the x-direction is presented in (b). }
    \label{fig: transport}
\end{figure}

The small MAE per unit cell reflects an isrotropic magnetic easy axis, enabling facile spin reorientation, which is highly favorable for both experimental detection and device applications. However, we found that the AHE exhibits strong dependence on spin orientation in rutile structures. Fig.~\ref{fig: transport}(a) shows the band structures of this superlattice with the \(\text{N\'eel}\) vector along x- and z-direction in the presence of SOC. The broken $\mathcal{PT}$ symmetry in the present strain-induced altermagnetic phase with SOC, giving rise to a finite AHE, whereas the preserved $\mathcal{PT}$ symmetry in the AFM phase enforces its vanishing. Reorienting the \(\text{N\'eel}\) vector redistributes Berry curvature across the Brillouin zone, allowing for a finite AHE. The analysis of AHE in the presence of SOC requires the use of magnetic space groups. When the \(\text{N\'eel}\) vector is aligned along [001], the magnetic space group is $P4_2'/mnm'$, which enforces vanishing AHE in all tensor components. In contrast, when the Néel vector is oriented along [100], the magnetic group is $Pnn'm'$, under which only the $\sigma_{zx}$ is permitted (\(\sigma_{zx} \sim 1400 S/cm\)), while both $\sigma_{xy}=\sigma_{yz}=0$ according to the symmetry, as shown in the Fig.~\ref{fig: transport}(b). This result is consistent with the symmetry properties of the Berry curvature, \(\mathcal{T}C_{2z}\Omega_x(k)=-\Omega_x(-k)\). Then we have \(\int dk_z\Omega_z(k)=\int dk_x \Omega_x(k)=0 \), which means \(\sigma_{xy}=\sigma_{yz}=0\) and \(\sigma_{zx} \neq0\). This finite intrinsic AHE in the $(\mathrm{CrO}_2)_1/(\mathrm{TaO}_2)_2$ superlattice is a hallmark of the AFM-to-altermagnetic transition, as seen in Fig.~\ref{fig: transport}(b). In RuO$_2$, such reorientation required large magnetic fields \cite{doi:10.1126/sciadv.aaz8809, 10.1038/s41928-022-00866-z}. In contrast, the minimal MAE suggests the same effect could be realized under much weaker fields, making the orientation-dependent AHE more experimentally accessible. In addition, the magnitude of the AHE is comparable to that observed in typical ferromagnets ($\sim 1000 S/cm$) \cite{s41567-018-0234-5}, highlighting the potential of this altermagnetic superlattice for hall-based spintronic applications. \par

We now turn to the topological phase transition, which occurs simultaneously with the magnetic transition. Without external strain, the intrinsic $(\mathrm{CrO}_2)_1/(\mathrm{TaO}_2)_2$ superlattice is a trivial AFM insulator (shown in the Supplementary Material). The strain-driven reorientation of the \(\text{N\'eel}\) vector alters the symmetry of the system, which in turn gives rise to the following observed topological phenomena. In the absence of SOC, the spin and orbital degrees of freedom are decoupled, allowing the crystal symmetry to be treated independently of spin. As shown in Fig.~\ref{fig: electronic structure}(b), the spin-up band structure exhibits two crossing points along the $\Gamma$–M direction and two additional crossings along the A–Z direction, all within the energy range of [–0.3, 0] eV relative to the Fermi level. Additionally, these crossings are points on the nodal rings in the $k_z=0$ and $k_z=0.5$ planes, as illustrated in Fig.~\ref{fig:AHE}(b), (c), and (e). Without SOC, the Hamiltonian commutes with M$_z$, allowing eigenstates to be labeled by mirror eigenvalues, which prevents hybridization between crossing bands and stabilizes the nodal rings in the spin-up channel. As a result, the nodal rings in the $k_z=0$ plane are protected solely by mirror reflection M$_z$, while those in the $k_z=\pm0.5$ planes are stabilized by the combined action of M$_z$ and the inversion symmetry. Besides, the gray nodal lines in Fig.~\ref{fig:AHE}(e) are protected by the twofold rotational symmetry $C_2$ along the [110] direction. Without this $C_2$ symmetry, they would evolve into a cylindrical band-crossing surface connecting the nodal rings in the $k_z=\pm 0.5$ planes. Due to time-reversal symmetry $\mathcal{T}$ and preserved fourfold rotational symmetry C$_{4z}$, the spin-down channel hosts identical nodal rings, resulting in a pair of doubly degenerate mirror-protected Weyl nodal rings.\par

Under the consideration of SOC with the \(\text{N\'eel vector}\) vector oriented along the [100] direction. The mirror-protected nodal rings are gapped in the Fig.~\ref{fig:AHE}(b) due to the spin-up will hybrids with the spin-down channel. Fig.~\ref{fig:AHE}(b) shows the band structure with and without SOC. This behavior contrasts with the robust Weyl nodal lines reported in rutile systems with out-of-plane magnetic moments (along the z-direction) \cite{PhysRevLett.134.096703}, where mirror and rotational symmetries jointly protect the nodal features even in the presence of SOC. The superlattice retains inversion symmetry even in the presence of SOC, which dictates that Weyl points must appear in pairs. We identified eight pairs of Weyl points using WannierTools \cite{WU2017} and a symmetrized tight-binding Hamiltonian constructed from maximally localized Wannier functions. The locations and chiralities of these Weyl points ($\chi=$ +1 and -1 shown in red and blue, respectively) are illustrated in Fig.~\ref{fig:AHE}(d). These Weyl points act as topological monopoles, sources and sinks of Berry curvature, symmetrically located at the $k_z=0$ and $k_z=\pm 0.5$ planes. The coordinates of the Weyl points with chirality ($\chi =+1$) are listed in Table 1.\par
\begin{table}[h]
    \centering
    \caption{Coordinates of Weyl points in units of reciprocal lattice vectors $(b_1,b_2,b_3)$.}
    \label{tab:weyl_points}
    \begin{tabular}{c c c c}
        \toprule
        Weyl point & $(b_1,b_2,b_3)$\\
        \midrule
        WP1 & $(0.108,-0.400,-0.500)$ \\
        WP2 & $(-0.108,-0.400,0.500)$ \\
        WP3 & $(0.052,0.445,0.000)$   \\
        WP4 & $(-0.052,0.445,0.000)$   \\
        WP5 & $(0.363,-0.065,-0.500)$ \\
        WP6 & $(-0.363,-0.065,0.500)$\\
        WP7 & $(0.448,0.060,0.000)$   \\
        WP8 & $(-0.448,0.060,0.000)$ \\
        \bottomrule
    \end{tabular}
\end{table}

The 16 Weyl points are located within two planes in the $k_z=0$ and $k_z=\pm 0.5$ planes. These nodes are stabilized by the preserved mirror symmetry $M_z$, which protects the linear crossings of bands with opposite mirror eigenvalues. Such symmetry-based placement of Weyl nodes is exceptional among most known Weyl semimetals. For example, in TaAs \cite{ncomms8373} and MoTe$_2$ \cite{PhysRevB.92.161107}, only a subset of Weyl points lie in mirror-symmetric planes, while others are located at generic k points. In contrast, the $(\mathrm{CrO}_2)_1/(\mathrm{TaO}_2)_2$ superlattice hosts all Weyl nodes constrained by mirror symmetry, reflecting the unique interplay between altermagnetism, crystal symmetry, and topology. This symmetry-enforced planar distribution of Weyl nodes enhances the robustness of their topological features and may simplify their experimental detection. Representative band dispersion along momentum-space cuts intersecting selected Weyl points is provided in the Supplementary Material. The presence of symmetry-protected Weyl nodes near the Fermi level suggests that their signatures—such as surface Fermi arcs or chiral transport—may be experimentally accessible via angle-resolved photoemission spectroscopy (ARPES) or magnetotransport measurements.\par
Unlike magnetic phase transitions coupled to structural transformations \cite{PhysRevB.109.144421}, the strain-induced topological AFM-to-altermagnetic magnetic transition reported here occurs without structural change. Unlike intrinsic CrSb \cite{acs.nanolett.5c00482}, this topological altermagnetic phase is tunable through external stress, providing an experimentally accessible platform. In contrast to the biaxial strain employed in 2D altermagnets \cite{2507.22474}, the uniaxial strain used here is easier to implement. Thus, the $(\mathrm{CrO}_2)_1/(\mathrm{TaO}_2)_2$ superlattice proposed here bridges altermagnetism and topology, opening a promising route for the design of strain-tunable spintronic systems and dissipationless electronics.\par
In summary, we have revealed that a modest uniaxial strain of 0.5$\%$ along the $c$ axis induces an AFM-to-altermagnetic phase transition in the $(\mathrm{CrO}_2)_1/(\mathrm{TaO}_2)_2$ superlattice. This transition is driven by interfacial charge transfer and is accompanied by the emergence of symmetry-protected topological features, including Weyl points and nodal lines. Our findings establish a viable strategy for engineering coexisting altermagnetism and topology in real materials, offering a tunable candidate for exploring novel spintronic and topological phenomena. In finalizing this work, we are aware of two related preprints that also explore aspects of topological altermagnetism in distinct material contexts \cite{2507.23173, 2507.22474}. Our work was developed independently and focuses on strain-tunable magnetic and topological transitions in the $(\mathrm{CrO}_2)_1/(\mathrm{TaO}_2)_2$ superlattice.\par
Dr. Wanfei Shan thanks Prof. Weidong Luo and Dr. Chengyang Xu for fruitful discussions. This work was supported by the U.S. Department of Energy, Office of Science, Basic Energy Sciences, Materials Sciences and Engineering Division. We also acknowledge Grant Numbers GBMF8048 and GBMF12976 from the Gordon and Betty Moore Foundation. Calculations were performed using the resources of the National Energy Research Scientific Computing Center, a DOE Office of Science User Facility supported by the Office of Science of the U.S. Department of Energy, and the Hoffman2 Shared Cluster provided by the UCLA Office of Advanced Research Computing’s Research Technology Group.\par

\clearpage              
\onecolumngrid          
\beginsupplement        

\begin{center}
\textbf{\large Supplemental Material for\\ ``Strain-Induced Antiferromagnetic-to-Altermagnetic Phase Transition and Topology in $(\mathrm{CrO}_2)_1/(\mathrm{TaO}_2)_2$ Superlattice''}
\end{center}

\section{Methods}
\begin{figure*}
    \centering
    \includegraphics[width=1\linewidth]{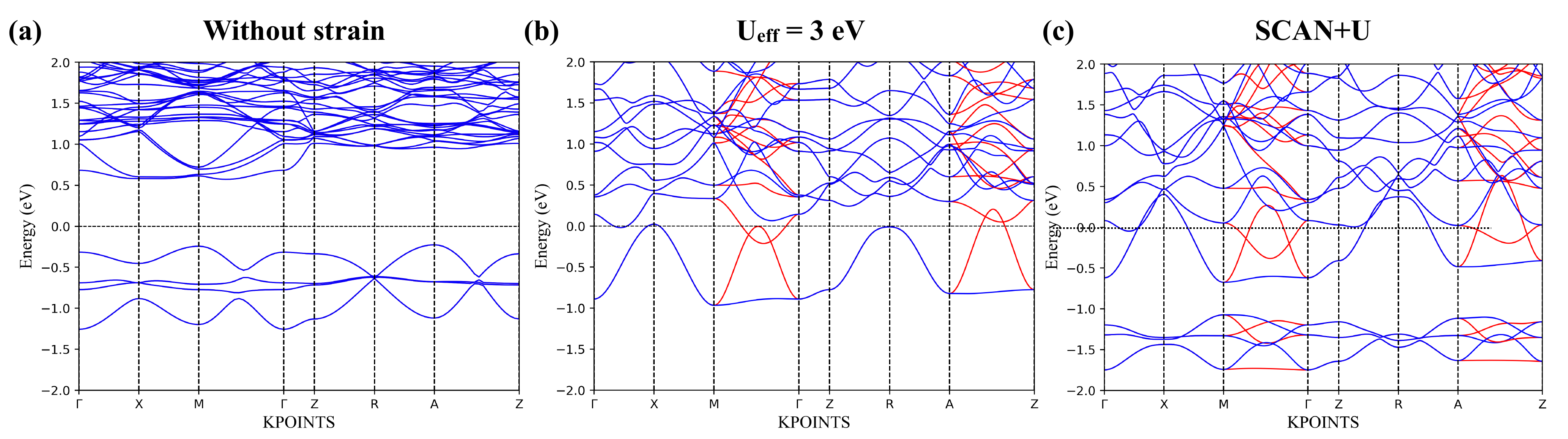}
    \caption{The band structures of the the (CrO$_2$)$_1$/(TaO$_2$)$_2$ Superlattice. The band structure of the intrinsic superlattice is shown in (a). The GGA+U calculation with U$_{eff}$= 3 eV is shown in (b). The SCAN+U (U$_{eff}$ = 0.2 eV) is shown in (c). The Fermi level is set to 0 eV.}
    \label{fig: different Hubbard U}
\end{figure*}

All density functional theory (DFT) \cite{PhysRev.136.B864,PhysRev.140.A1133} calculations are performed with the help of the Vienna Ab initio Simulation Package (VASP) \cite{PhysRevB.54.11169} using the projector augmented wave (PAW) \cite{PhysRevB.50.17953} method. The exchange-correlation functions is chosen as the generalized gradient approximation plus Hubbard U (GGA+U) method \cite{PhysRevB.44.943, PhysRevLett.77.3865}. We employ a Hubbard U$_{eff}$ of 2, 3 eV  within the GGA+U calcuations \cite{10.1063/1.481183, 10.1088/1361-648X/aca19a, Shan_2023} for the (CrO$_2$)$_1$/(TaO$_2$)$_2$ superlattice. We first perform calculations for the intrinsic (CrO$_2$)$_1$/(TaO$_2$)$_2$ superlattice, and obtain a band structure that is consistent with the main text results, as shown in Fig.~\ref{fig: different Hubbard U}(a).Subsequently, we carried out calculations incorporating an external uniaxial strain along the $c$-direction. A $12\times12\times12$ kmesh, the energy cutoff of 400 eV, and an energy convergence criterion of 10$^{-6}$ were employed. We performed calculations using the strongly constrained and appropriately normed (SCAN) meta-generalized gradient approximation (meta-GGA) functional \cite{PhysRevLett.115.036402, PhysRevMaterials.4.045401}, supplemented by a small Hubbard correction of U$_{eff}$ = 0.2 eV. The resulting band structure near the Fermi level closely matches that from the GGA+U calculations with $U=2eV$, confirming the stability of the topological and magnetic characteristics against the choice of exchange-correlation functional. The maximally localized Wannier functions were constructed by s and d orbitals of Cr, d orbitals of Ta, and p orbitals of O with Wannier90 \cite{MOSTOFI20142309} software, and then the Hamiltonian was symmetrized by wannhr\_symm\_Mag \cite{yue2025wannhr} before any analysis of topological properties.\par
The result in fig. \ref{fig: different Hubbard U} (b) and (c) exhibit the same topological magnetic phase transition as what is described in the main content with GGA+U (U$_{eff}$= 2 eV). The band structure in fig. \ref{fig: different Hubbard U} (b) and (c) refer to the results with U$_{eff}$ = 3 eV and the SCAN+U result, respectively. So far, the magnetic moments in these calculations are all larger than 3$\mu B$, e.g. 3.3 $\mu B$ with U$_{eff}$ = 2 eV, 3.4 $\mu B$ with U$_{eff}$ = 3 eV, and 3.2 $\mu B$ with SCAN+U. So, all these calculations address the same topological magnetic phase transition in the superlattice.\par
\section{Charge transfer in the (CrO$_2$)$_1$/(TaO$_2$)$_2$ Superlattice}
\begin{figure}
    \centering
    \includegraphics[width=1\linewidth]{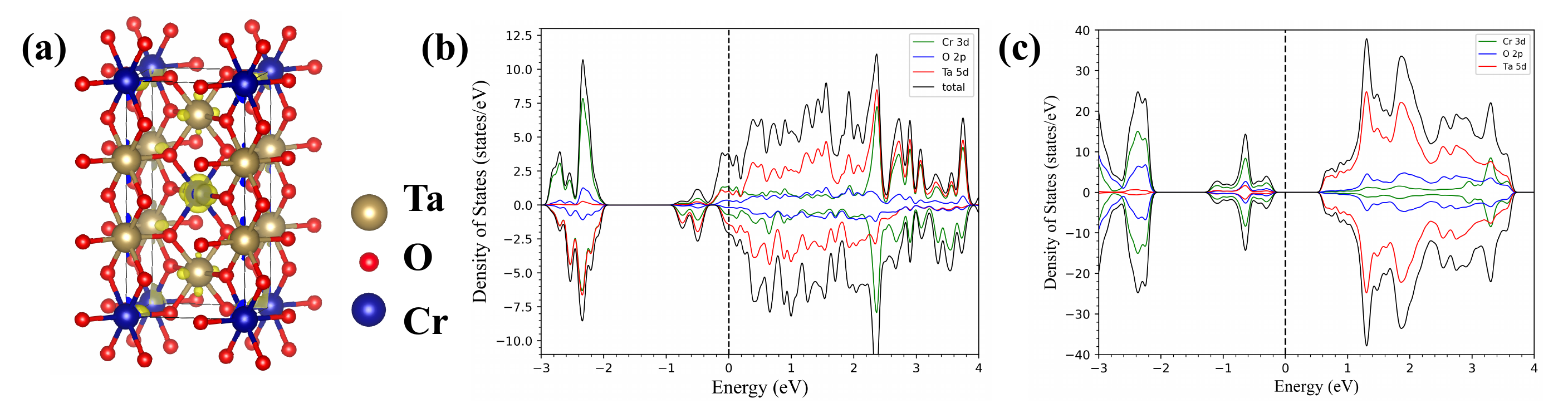}
    \caption{Partial charge density, and the projected DOS of the (CrO$_2$)$_1$/(TaO$_2$)$_2$ superlattice with and without external strain. (a) Isosurface plot of the partial charge density integrated within the energy window from $–1$ eV up to the Fermi level. Blue, red, and dark yellow spheres denote Cr, O, and Ta atoms, respectively. (b) and (c) Projected DOS of the (CrO$_2$)$_1$/(TaO$_2$)$_2$ superlattice; the Fermi level is set to 0 eV.}
    \label{fig: PDOS}
\end{figure}

The charge transfer could be exhibited especially from the projected density of states (PDOS) in fig. \ref{fig: PDOS}. By comparing the PDOS of the superlattice with and without applied strain, as shown in Fig.\ref{fig: PDOS}(b) and (c), a clear signature of charge transfer emerges, primarily involving Cr 3$d$ orbitals and, to a lesser extent, Ta 5$d$ orbitals. Although four Ta atoms contribute to the relevant energy range in the superlattice, the PDOS per Ta atom remains significantly lower than that of Cr, as also seen in Fig.\ref{fig: PDOS}(a).\par

\section{Doping case}
\begin{figure}
    \centering
    \includegraphics[width=1\linewidth]{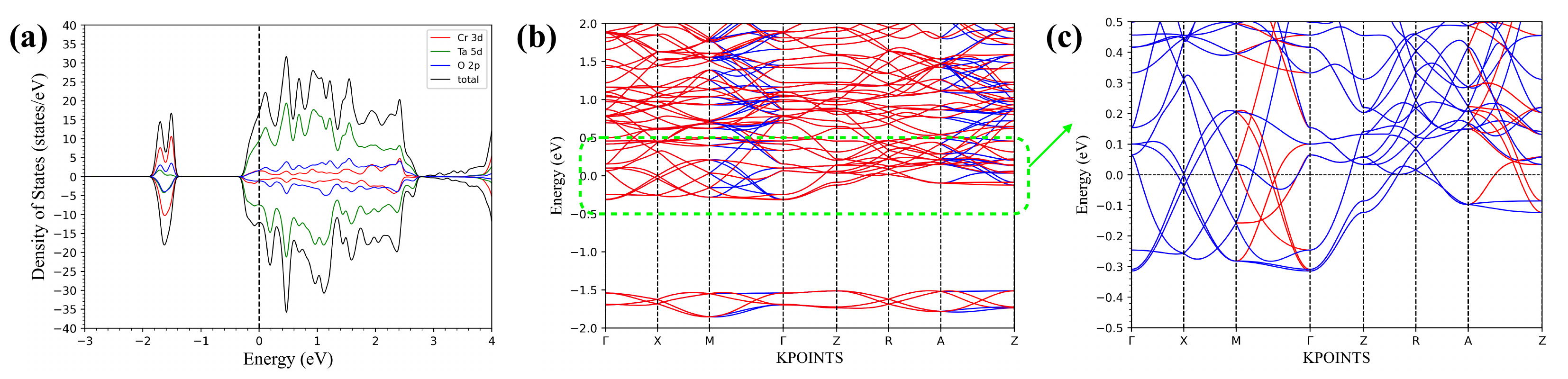}
    \caption{Electronic properties of the doping (CrO$_2$)$_1$/(TaO$_2$)$_2$ superlattice. (a) and (b) show the PDOS and band structure of the doping (CrO$_2$)$_1$/(TaO$_2$)$_2$ superlattice, respectively. (c) describes the enlarged energy range of [-0.5, 0.5] eV with respect to the Fermi level. The Fermi level is set to 0 eV in all these figures.}
    \label{fig: doping}
\end{figure}

To further verify the mechanism behind the phase transition, we perform an additional calculation by introducing a small amount of electron doping into the (CrO$_2$)$_1$/(TaO$_2$)$_2$ unit cell, such that the excess charge per Cr atom remains significantly below one electron. The PDOS and band structure are shown in fig. \ref{fig: doping}. The resulting band structure indicates that the altermagnetic ground state is robust, with comparable band splittings preserved along the $\Gamma$–$M$ and $A$–$Z$ high-symmetry directions.\par
In the doped case, the bands near the Fermi level exhibit a high density of hybridized crossings, in marked contrast to the relatively clean dispersion obtained under applied strain in the main text. This spectral complexity would hinder unambiguous experimental identification of the topological features. Accordingly, we confine the detailed discussion in the main text to the strain-tuned system. Nevertheless, the doping results already uncover the decisive influence of charge transfer on the phase transition. Besides, it inspires people with a method to tune the state of this superlattice by either external strain or doping.\par

\section{Magnetic exchange interaction}
\begin{figure}
    \centering
    \includegraphics[width=1\linewidth]{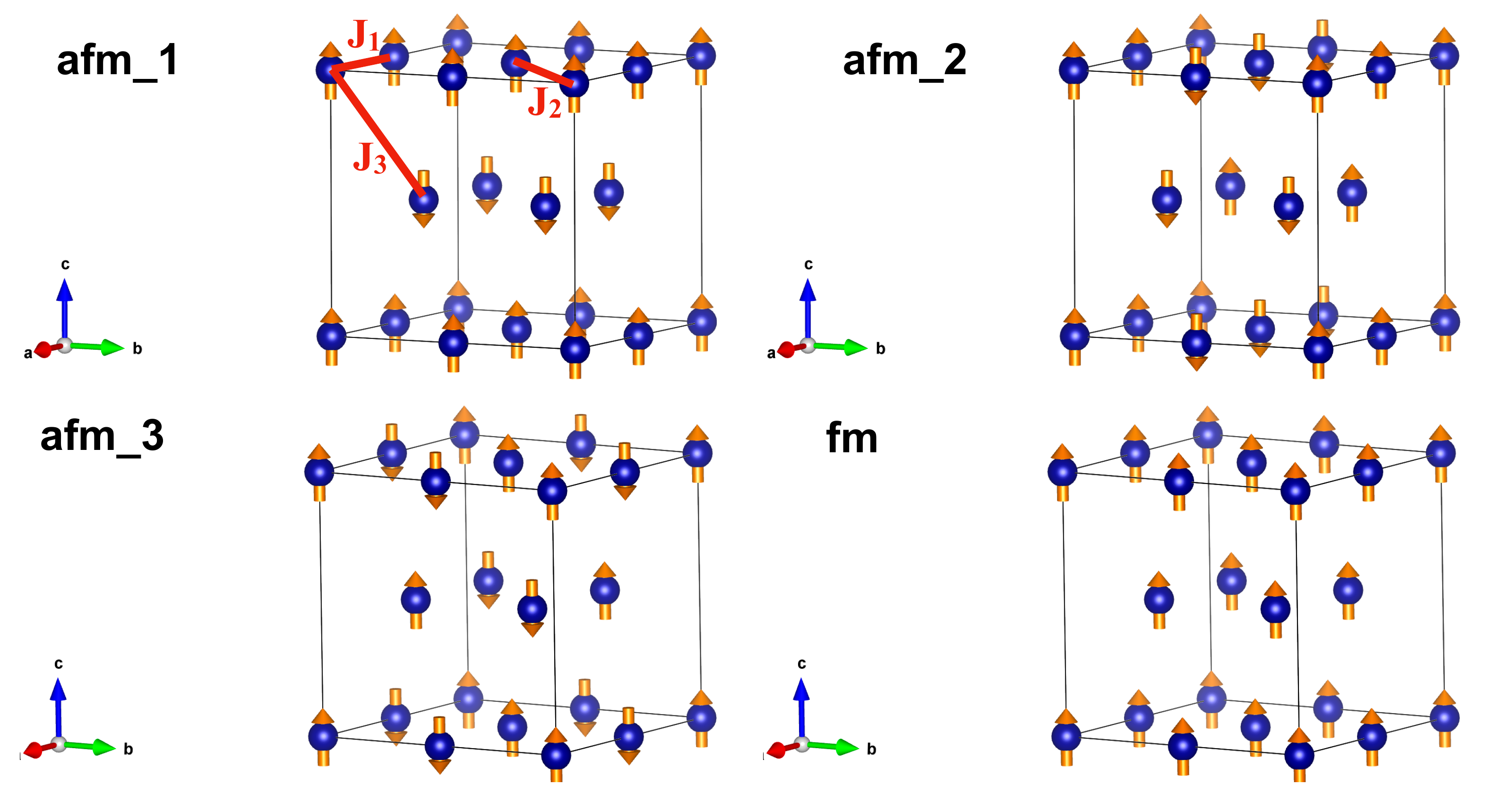}
    \caption{Different magnetic configurations of the (CrO$_2$)$_1$/(TaO$_2$)$_2$ superlattice are shown in the figure with only Cr atoms left in the $2\times2\times1$ supercell.}
    \label{fig: magnetic configurations}
\end{figure}

We calculate nine different magnetic configurations, including eight AFM configurations and FM. By mapping the total energies of these configurations onto a semi-classical Heisenberg model, we construct a set of four independent equations involving three exchange parameters. The spin arrangements associated with these equations are illustrated in Fig.~\ref{fig: magnetic configurations}, where only Cr atoms are displayed within the $2\times2\times1$ supercell. The explicit mapping between the magnetic configurations and exchange interactions is summarized in:

\begin{align*}
    4J_1+4J_2-8J_3+E_0=E_{afm1}\\
    0-4J_2+0+E_0=E_{afm2}\\
    -4J_1+4J_2+0+E_0=E_{afm3}\\
    4J_1+4J_2+8J_3+E_0=E_{fm}
\end{align*}
with $J_1$, $J_2$, and $J_3$ describe the exchange interactions in the superlattice.

\section{Compared with the former cases}
RuO$_2$, the prototypical rutile altermagnet and the first material predicted to host an altermagnetic phase, has shown little evidence of spin splitting in recent spin- and angle-resolved photoemission spectroscopy (SARPES) experiments, in agreement with nonmagnetic band structure calculations \cite{PhysRevLett.133.176401, PhysRevLett.132.166702}. This discrepancy has been attributed to the use of a relatively large Hubbard $U$ ($\geq$ 2 eV) in prior theoretical studies, which potentially overestimates both the magnetic moment and spin splitting. Meanwhile, calculations performed with a smaller Hubbard $U$ (0–1 eV) yield results consistent with experimental observations. These results uncover the importance of Hubbard $U$ in determining the magnetic properties of $d$ orbital systems. Our results, by comparison, demonstrate that the altermagnetic state in this superlattice persists and has strong magnetic moments even with the smallest applied value of $U_{\mathrm{_eff}} = 2$ eV, the lowest physically reasonable estimate for Cr-based compounds. This robustness highlights its viability as a robust rutile altermagnet. \par
The energy splitting related by \{$C_{4z}|\tau$\} translation between the two Cr atoms in the superlattice only happens along $\Gamma$ to $M$ and $A$ to $Z$ directions in the band structure in the main text. The largest energy split is the same magnitude as RuO$_2$ \cite{PhysRevX.12.031042}. It reflects the relatively strong exchange interaction between two sublattices, which is consistent with the meV magnitude of the exchange interaction in the superlattice. The exchange interaction here is substantially larger than that in MnF$_2$. In marked contrast, spin splitting in MnF$_2$ originates from weak higher-order anisotropic exchange interactions, which are nearly negligible. As a result, MnF$_2$ was verified by inelastic neutron scattering (INS) measurements to behave effectively as a conventional \(\text{N\'eel}\) antiferromagnetic \cite{PhysRevLett.134.226702}. This comparison supports identifying the $(\mathrm{CrO}_2)_1/(\mathrm{TaO}_2)_2$ superlattice as a promising rutile altermagnet candidate.\par
\section{The Nodal line}
\begin{figure}
    \centering
    \includegraphics[width=1\linewidth]{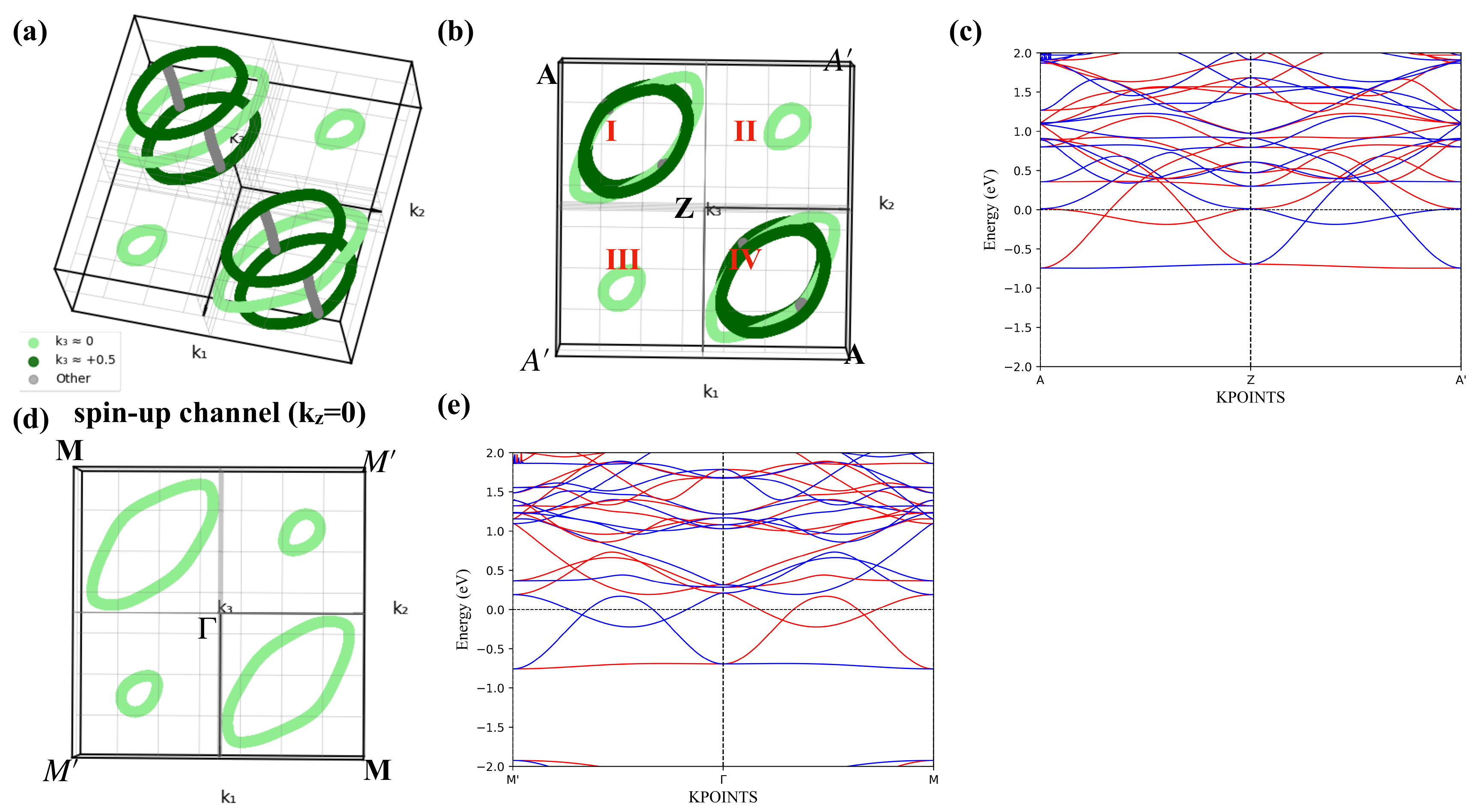}
    \caption{Nodal-line distribution for the spin-up channel in the Brillouin zone of the superlattice is shown in (a). The top view of the nodal lines, projected onto the $k_x$–$k_y$ plane, is shown in (b). The nodal rings in the $k_z$=0 plane are shown in (d). The band structure along the A-Z-A' direction is shown in (c). The band structure along the $M'-\Gamma-M$ direction is shown in (e).}
    \label{fig: Nodal line}
\end{figure}

In the absence of spin-orbit coupling, the spin channels are decoupled, allowing us to analyze the spin-up band structure independently. The nodal lines in the spin-down channel are symmetry-related via the space group operations, such as the fourfold rotation $C_{4z}$. As a consequence, the nodal rings in the spin-down channel are symmetry-equivalent to those in the spin-up channel and form an identical network of nodal structures in momentum space. That is consistent with the band structure in fig. \ref{fig: Nodal line}(c) and (e). These nodal rings are robust against weak perturbations that preserve the underlying mirror and rotational symmetries. \par

\section{The Weyl points}
\begin{figure}
    \centering
    \includegraphics[width=1\linewidth]{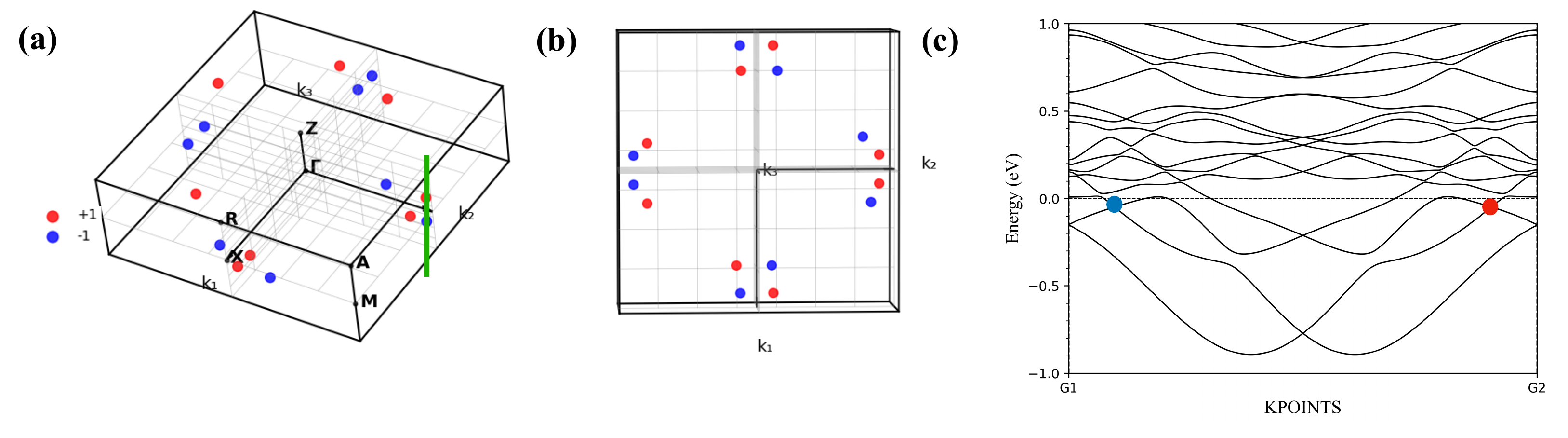}
    \caption{Weyl points of the (CrO$_2$)$_1$/(TaO$_2$)$_2$ superlattice in the Brillouin zone are shown in (a) and top view in (b), respectively. (c) describes the band structure along one pair of Weyl points we picked up in (a).}
    \label{fig: Weyl points}
\end{figure}

On considering SOC, eight pairs of Weyl points emerge in the Brillouin zone of the superlattice, as shown in fig. \ref{fig: Weyl points}(a) and (b). To verify the linear band crossings associated with these Weyl points, we perform a local band structure calculation along high-symmetry directions around one representative pair, as presented in fig \ref{fig: Weyl points}(c). The band structure in fig \ref{fig: Weyl points}(c) confirms a clear linear crossing between the conduction and valence bands at the predicted Weyl nodes.

\section{Experiment Realization}
Given recent advances in oxide thin film synthesis, the fabrication of the (CrO$_2$)$_1$/(TaO$_2$)$_2$ superlattice is experimentally feasible. Both CrO$_2$ and TaO$_2$ can crystallize in the rutile phase (as mentioned in the main text), with in-plane lattice constants of approximately 4.42 Å and 4.98 Å, respectively, leading to an in-plane lattice mismatch of about 7$\%$. For comparison, rutile TiO$_2$ has been widely employed as a substrate for high-quality epitaxial CrO$_2$ thin films despite a similar lattice mismatch of $\sim4\%$, as demonstrated in numerous studies \cite{Ventrice_2007, 10.1063/1.1736326,10.1063/1.2188045}. In addition, thin films of rutile NbO$_2$ epitaxially grown on rutile TiO$_2$ (100) or Al$_2$O$_3$ (0001) substrates have been successfully used as a template to stabilize rutile TaO$_2$ under appropriate growth conditions \cite{MURAOKA2016125}. Therefore, by employing an appropriate buffer layer (e.g., TiO$_2$) and leveraging high-precision deposition techniques such as Pulsed Laser Deposition (PLD) or Molecular Beam Epitaxy (MBE) with in-situ monitoring, the proposed (CrO$_2$)$_1$/(TaO$_2$)$_2$ superlattice can be experimentally realized with current state-of-the-art methods.\par
In practice, such small strain values (0.5$\%$ along c-direction) can be achieved either via coherent epitaxial strain imposed by a lattice-mismatched substrate or through post-growth techniques such as piezoelectric-based strain application or thermal strain engineering. Besides, ultrafast optical experiments could also apply such small strain on the lattice of complex oxide heterostructures via a vibrational mode \cite{PhysRevLett.108.136801}. Therefore, the moderate strain magnitude considered in our work is both realistic and experimentally accessible.\par

\end{document}